\documentclass[aps,prd,twocolumn,groupedaddress,nofootinbib,longbibliography,superscriptaddress]{revtex4-2}

\usepackage{amsfonts,amsmath,amssymb}
\usepackage{physics}

\usepackage{mathrsfs}
\usepackage{nicefrac}
\usepackage{graphicx}
\usepackage{color}
\usepackage{soul} 
\usepackage{hyperref}
\usepackage{accents}
\usepackage{subfigure}
\usepackage{soul}
\usepackage[normalem]{ulem}
\usepackage{hyperref}
\usepackage{xcolor}
\usepackage{fleqn}
\usepackage{todonotes}

\hypersetup{
    colorlinks=true, 
    linktoc=page,    
    linkcolor=blue,  
    urlcolor=magenta
}

\begin{document}
\title{Spectrogram correlated stacking: A novel time-frequency domain analysis of the Stochastic Gravitational Wave Background }


\author{Ramit Dey}
\email[]{rdey5@uwo.ca}
\affiliation{Department of Physics and Astronomy, The University of Western Ontario, London, ON N6A 3K7, Canada}
\affiliation{Perimeter Institute For Theoretical Physics, 31 Caroline St N, Waterloo, Canada}

\author{Lu\'{i}s Felipe Longo Micchi}
\email[]{luis.longo@ufabc.edu.br}
\affiliation{
Center for Natural and Human Sciences, Universidade Federal do ABC, Santo Andr\'{e}, SP 09210-170, Brazil}
\affiliation{Institute for Gravitation \& the Cosmos, The Pennsylvania State University, University Park, PA 16802, USA}
\affiliation{Department of Physics, The Pennsyvlania State University, University Park, PA 16802, USA}
\affiliation{Theoretisch-Physikalisches Institut, Friedrich-Schiller-Universit\"at Jena, 07743, Jena, Germany}

\author{Suvodip Mukherjee}
\email[]{suvodip@tifr.res.in}
\affiliation{Department of Astronomy \& Astrophysics, Tata Institute of Fundamental Research, 1, Homi Bhabha Road, Colaba, Mumbai 400005, India}

\author{Niayesh Afshordi}
\email[]{nafshordi@pitp.ca }
\affiliation{Waterloo Centre for Astrophysics, University of Waterloo, Waterloo, ON, N2L 3G1, Canada}
\affiliation{Department of Physics and Astronomy, University of Waterloo,
200 University Ave W, N2L 3G1, Waterloo, Canada
}
\affiliation{Perimeter Institute For Theoretical Physics, 31 Caroline St N, Waterloo, Canada}

\begin{abstract}
The astrophysical stochastic gravitational wave background (SGWB) originates from numerous faint sub-threshold  gravitational wave (GW) signals arising from the coalescing binary compact objects. This background is expected to be discovered from the current (or next-generation) network of GW detectors by cross-correlating the signal between multiple pairs of GW detectors. However, detecting this signal is challenging and the correlation is only detectable at low frequencies due to the arrival time delay between different detectors. In this work, we propose a novel technique, \texttt{Spectrogram Correlated Stacking} (or \texttt{SpeCS}), which goes beyond the usual cross-correlation (and to higher frequencies)  by exploiting the higher-order statistics in the time-frequency domain which accounts for the \textit{chirping} nature of the individual events that comprise SGWB. 
We show that \texttt{SpeCS} improves the signal-to-noise for the detection of SGWB by a factor close to $8$, compared to standard optimal cross-correlation methods which are tuned to measure only the power spectrum of the SGWB signal.
\texttt{SpeCS} can probe beyond the power spectrum and its application to the GW data available from the current and next-generation GW detectors would speed up the SGWB discovery.

\end{abstract}

\maketitle

\section{Introduction}

The direct detection of gravitational waves (GWs) by the LIGO/Virgo/KAGRA (LVK) collaboration has been sourced by individual coalescences of compact binaries for which the signal shows up as coherent and resolved waveforms in the detectors  \cite{LIGOScientific:2016aoc,LIGOScientific:2020ibl,LIGOScientific:2021psn}.  
Based on the current event rate estimations, these loud events only form a tiny fraction of the gravitational-wave sky (e.g., \cite{Antonini:2020xnd,KAGRA:2021kbb}). It is expected that numerous unresolved stellar-mass binary mergers would add up incoherently and lead to the production of a stochastic gravitational wave background (SGWB; e.g., \cite{Allen:1997ad,Regimbau_2011,Wu:2011ac,Zhu:2011bd,Romano:2016dpx,Christensen:2018iqi,Renzini:2022alw}). While some backgrounds, such as the one due to inflationary tensor modes are generated as a (nearly-gaussian quantum) stochastic process \cite{1979JETPL..30..682S,PhysRevD.55.R435,Guzzetti:2016mkm}, others (e.g., due to unresolved astrophysical sources such as black hole binaries \cite{Regimbau:2008nj,Rosado:2011kv,Zhu:2011bd}) would appear stochastic due to the observational limitation of current detectors to identify individual merger events. For the astrophysical SGWB of the second kind, it is by definition at the threshold of detection and thus, one can interpret it as a detector-dependent observable \cite{Cornish:2015pda}. 

 The conventional approach to detect the SGWB consists of performing a cross-correlation based analysis on the strain measurements obtained by a pair of GW detectors \cite{Allen:1997ad,Mitra:2007mc,LIGOScientific:2014gqo,Christensen:2018iqi, vanRemortel:2022fkb}. Based on this technique, the LVK collaboration has placed upper limits on the strength of the isotropic and anisotropic SGWBs in the frequency range 20-1726 Hz \cite{KAGRA:2021rmt,KAGRA:2021mth,KAGRA:2021kbb} using the latest third observation (O3) run, limiting the dimensionless energy density to $\Omega_{\rm{GW}}(f) \leq 3.4\times 10^{-10}$ at 25 Hz for a GW background signal of the form $\Omega_{\rm GW}(f)= f^\alpha$ with a spectral index of $\alpha=2/3$.


Though this measurement, based on the strain (cross-) power spectrum, has obtained the most stringent bounds on the SGWB to date, one of its key aspects which remains unexplored (at least in real data) is the temporal fluctuation of the SGWB power due to non-overlapping nature (or low duty cycle)  of the merger events that comprise the signal \cite{Mukherjee:2019oma, PhysRevD.104.063518}. 
In  other words, the low duty cycle results in a non-Gaussian stochastic background, which can be probed using higher-order statistics. 
Several methods have been proposed to study the non-Gaussian aspect of the SGWB \cite{Mukherjee:2019oma, PhysRevD.104.063518, Smith:2017vfk, Drasco:2002yd, Ballelli:2022bli,Martellini:2014xia,2023PhRvD.107f3027B, Braglia:2022icu}.
Higher order detection statistics (e.g., as proposed in \cite{Seto:2009ju,Martellini:2015mfr}) can trace/extract the excess information present in the SGWB `popcorn' signal and show noticeable sensitivity gain when the signal has intrinsic non-gaussianity.

In this paper, we introduce a novel time-frequency domain (4th order) statistical estimator which can capture the time-dependent SGWB signal. This new method of \textit{Spectrogram Correlated Stacking} (\texttt{SpeCS}) is based on performing a cross-correlation of the spectrograms, which is obtained by performing a short-time-Fourier-transform on the GW strain as measured by different detectors. 
In order to test various features of this detection statistics, we simulate two sets of SGWBs by injecting individual low amplitude (low Signal-to-noise-ratio) binary merger events into synthetic LIGO noise (generated using the PSD of the O3 run for LVK detectors \cite{LIGOScientific:2023vdi}) and demonstrate the performance of this method for different merger rates of binary objects.

The paper is organized as follows: The key motivation behind this method is described in Sec. \ref{motivation} and the formalism of the technique \texttt{SpeCS} is shown in Sec. \ref{time-freq}. In Sec. \ref{sims}, we show the application of the technique on the simulated GW time series and its measurability from a network of GW detectors. We also contrast the \texttt{SpeCS} performance with the standard cross-correlation method. 
  Finally, in Sec. \ref{conc} we discuss the conclusion and the future outlook.




\section{Motivation behind spectrogram search of SGWB}\label{motivation}
\begin{figure}
    \centering
    \includegraphics[width=0.6\textwidth]{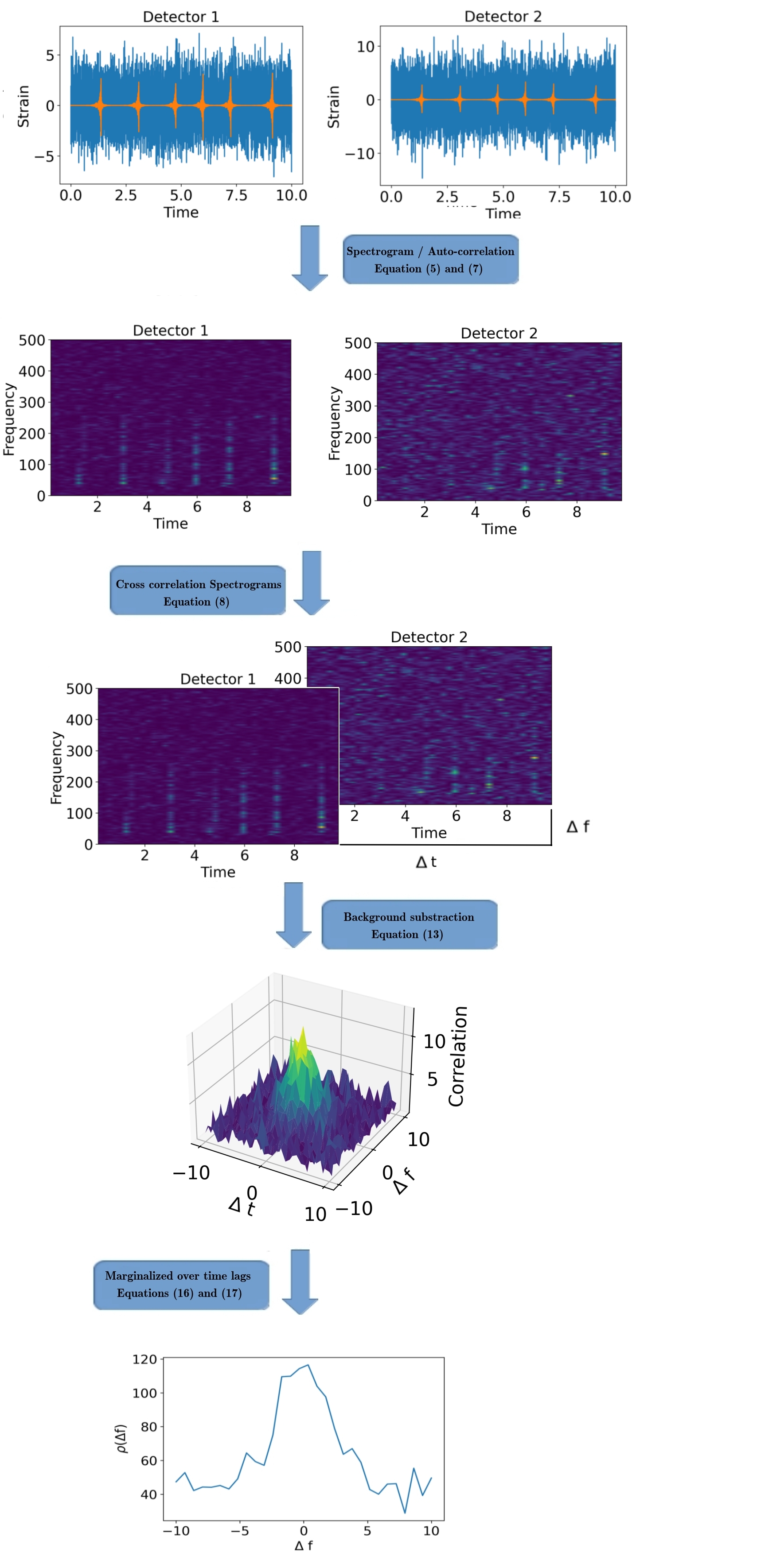}
    \caption{Flow chart of the proposed \texttt{SpecCS} method. We note that during a real application of the method, the signals of interest are not expected to be individually resolvable. The overall amplitudes here are exaggerated for visualization purposes. Axes are in arbitrary units.}
    \label{fig:my_label}
\end{figure}

    The astrophysical SGWB is observable in the high-frequency band ($f \geq 20$ Hz) of GW which is detectable from the ground-based network of GW detectors such as LVK\cite{LIGOScientific:2014pky, VIRGO:2014yos} and in the future from the Einstein telescope \cite{Punturo:2010zz,Sathyaprakash:2011bh} and Cosmic Explorer \cite{Reitze:2019iox}. As the contribution from these sources is extremely weak in comparison to the detector noise, they are not detectable as individual events. One of the key physical aspects of the astrophysical SGWB signal detectable in the high-frequency band is that the GW signals are non-overlapping with each other, or in other words, the duty cycle $\mathcal{D}$ of the GW signal defined as \cite{Wu:2011ac, Rosado:2011kv} 

\begin{align}\label{duty-cyc}
    \begin{split}
\mathcal{D}(>f_{\rm min})\equiv \int_{f_{\rm min}} \dd f \int \dd{z} \dd{\cal M}_z\,  R(z,{\cal M}_z) \frac{\dd\tau_d}{\dd f},    \end{split}
\end{align}
is $< 1$ for frequencies above $f_{\rm min}\sim 20$ Hz with the merger rates feasible according to the latest measurements from GWTC-3 \cite{Wu:2011ac,PhysRevD.104.063518,KAGRA:2021kbb}. In the above equation, $R (z)$ is the merger rate as a function of the cosmological redshift $z$ and the term $\frac{\dd\tau_d}{\dd f}$ denotes the duration that the GW signal spends at frequency $f$ 
\begin{align}\label{tauf}
\frac{\dd\tau_d}{\dd f}({\cal M}_z,f)= \frac{5c^5}{96\pi^{8/3} G^{5/3} \mathcal{M}_z^{5/3} f^{11/3}},
\end{align}
where $\mathcal{M}_z$ is the redshifted chirp mass of the GW sources. For values of the duty cycle $\mathcal{D}(f>f_{\rm min}) <1$, the observed SGWB is dominated by the non-overlapping GW signal as the timescale between the two events $T_R(z) \sim \left[\int R(z,{\cal M}_z) d{\cal M}_z\right]^{-1}$ is large compared to the time the signal of each event spends above our minimum frequency, i.e. $\sim f_{\rm min} \frac{\dd\tau_d}{\dd f}$. In the opposite limit, if the rate of the events is very large for example, then $T_R(z)$ can become small and hence the sources contributing to the SGWB will be overlapping. The presence of a non-overlapping GW signal implies that every individual signal though hidden in noise will still be distinguishable from each other in the time domain. This produces an interesting time-dependent aspect in the analysis of SGWB signal which is the motivation behind developing \texttt{SpeCS}. As shown in Fig. \ref{fig:my_label}, the \texttt{SpeCS} method is based on stacking the cross-correlation of the spectrograms of the GW strain signals, as measured by two different detectors, which is naturally a function of the difference in time and frequencies of the two independent data streams, $\Delta t$ and $\Delta f$ \footnote{More sensitive detectors can see higher redshift and frequency ranges, eventually leading to overlapping signals. However, since \texttt{SpeCS} is quartic in the strain of the GW source, it will be dominated by the brightest sub-threshold sources that are likely to remain non-overlapping. }.

The presence of the non-overlapping signal implies that each GW source will preserve its unique \textit{chirp} behaviour without mixing it up with another GW signal at the same observation time and frequency. Though the GW signal is completely buried under the noise fluctuation of the detectors, it can still preserve the very intrinsic chirp behaviour. The chirp behaviour of the GW signal will lead to a coherent structure in the time-frequency domain, across the detectors, that will be distinguishable from the (uncorrelated) noise properties of individual detectors \cite{Mukherjee:2019oma, PhysRevD.104.063518}. 
Although this signal will be much weaker than the noise fluctuation, three aspects will make it possible to distinguish from detector noise, (i) the chirping property will be similar for every source in all the detectors, but will be shifted in time, depending on the time difference for the GW signal to reach two detectors, which will depend on the sky position of the GW sources, (ii) GW signal will show a correlation between different frequencies as a function of time, and (iii) the noise property of every detector will be different and won't have coinciding chirp behaviour every time GW signal is present in the data. 
These three aspects make it possible to isolate the signal from the noise. Though glitch-like noise can also produce chirp-like behaviour, it won't be happening every time a signal is present in the data of all the detectors. However, due to the weak nature of the individual signal in comparison to the noise, individual spectrograms are not measurable and one needs to combine multiple events over a large observation time over different pairs of detectors in order to detect them as SGWB.

One of the main advantages of a spectrogram search of the SGWB signal instead of the usual cross-correlation power spectrum searches comes from the fact that it takes into account the correlation between different frequencies of the GW signal. Such a combination of a signal at different frequencies can help in extracting more information about the GW signal which is beyond the power spectrum analysis. The sporadic and chirp nature of the GW signal leads to an intrinsic non-Gaussian time-dependent structure, which can be captured by the new method proposed in this work.  Furthermore, the spectrogram analysis can possibly also help in achieving a better understanding of the source parameters than only a power spectrum analysis.

\section {Time-frequency domain detection statistics}\label{time-freq}

In this section, we introduce the time-frequency domain spectrogram estimator, \texttt{SpeCS}, which can be applied for the search of SGWB signal beyond the currently existing technique \cite{Allen:1997ad,Romano:2016dpx,KAGRA:2021kbb}.  
 Let us  consider $N$ gravitational wave detectors whose data are given as dimensionless strain measurements $s_{k}(t)$, where $k=1,2\hdots N$ labels the detector. We can decompose this measurement in terms of the contribution from signal and noise as
\begin{align} \label{sig}
    s_{k}(t)=\hat h_k(t) +n_k(t) ,
\end{align}
where $\hat h_k(t)=\sum_i F^k_i(t)h_i(t)$ is the observed GW signal as seen by a detector, with $h_{i}(t)$ denoting the actual signal for the
polarization state ``$i$", and the corresponding detector antenna pattern
$F^k_i$(t). The noise in the detector is denoted by $n_{k}(t)$. 
This signal can be represented in the time-frequency domain by performing a discrete short-time-Fourier-transform transform as
\begin{align}\label{dsft}
W_{s_i}(t_m,f)=\sum_{n=-\infty}^{\infty} s_i(t_n)g(t_n-t_m)e^{2\pi if t_n},
\end{align}
where $W_{s_i}$ is the signal in the time-frequency domain,  $g(t_n-t_m)$ is a window function and $t_m$ is the discrete time. In practice, we use this particular transformation \eqref{dsft} in order to maintain uniform time resolution at different frequency bins.

The cross-spectrum at a finite time difference $\Delta t$ and finite frequency difference $\Delta f$ can be written as

\begin{align} \label{cross-power}
S_{s_i, s_j}(\Delta t,\Delta f)\equiv \sum_t \sum_f  W_{s_i}(t,f)W_{s_j}(t+\Delta t,f+\Delta f) .
\end{align}
The above expression for $\Delta f=0$ calculates the cross-correlation power spectrum of the signal estimated at some value of $\Delta t$. One can further construct an optimal estimator based on Eq. \eqref{cross-power} by using Weiner filtering. This would further maximize the SNR, as explored in previous studies \cite{Thrane:2013kb} but it is beyond the scope of this paper.


\subsection{Spectrogram correlated stacking}

The higher order correlation between a pair of detectors can be expressed in a similar way in terms of the spectrogram (auto-correlation signal) $S_{s_i}(t,f)$ defined as 
\begin{align}
S_{s_i}(t,f)=|W_{s_i}(t,f)|^2 .
\end{align}
Here, the variables $(t,f)$ can be either discrete or continuous depending on the method/algorithm used for mapping the signal into the time-frequency domain. 

For the detection and analysis of the SGWB signal, we define the new \texttt{SpeCS} statistic in terms of the spectrogram as
\begin{align} \label{speccs}
P_{cc} (\Delta t, \Delta f)\equiv\sum_t\sum^{f_{\rm max}}_{f=f_{\rm min}} S_{s_1}(t,f) S_{s_2}(t-\Delta t, f-\Delta f) \, ,
\end{align}

$\Delta t$ and $\Delta f$ denote the time and frequency lag correspondingly. 
This detection statistics $P_{cc}$ can be further decomposed in terms of the detector noise and SGWB contribution to the spectrogram as \footnote{For the spectrogram of the signal $S_{s_i}(t,f ) \neq S_{h_i}+S_{n_i}$, but in this case, while taking the cross-correlation of the spectrograms within the integral the given decomposition holds. This is true because the cross-term involving the noise and GW signal in the time-frequency domain becomes negligible while integrating over $t$.}
\begin{align} \label{scc}
&P_{cc} (\Delta t, \Delta f) =\sum_t\sum^{f_{\rm max}}_{f=f_{\rm min}} [S_{h_1}+S_{n_1}] [S_{h_2}+S_{n_2}]
\nonumber\\
&=\sum_t\sum^{f_{\rm max}}_{f=f_{\rm min}}[S_{h_1}S_{h_2} + S_{h_1}S_{n_2} + S_{n_1}S_{h_2} + S_{n_1}S_{n_2}] \,.
\end{align}
Since the individual pixels of the spectrogram denote the power of the signal at a given frequency and time, the cross-correlation between the spectrogram of the noise and signal (or noise from the two different detectors) would not be zero.

The quantity $P_{cc}(\Delta t, \Delta f)$ is not an unbiased estimation as it will include contamination from noise. 
However, the noise feature in the time-frequency domain will exhibit different features from the signal. The signal will exhibit a strong correlation primarily around $\Delta t=0$, and $\Delta f=0$. There will be a spread in the frequency domain due to the \textit{chirping} behaviour of the signal or noise. When combining different pairs of detectors, the signal will show up as a constructive power, which is more in comparison to the noise fluctuations. This is because the signal will be common between all the detectors, but noise won't be common. 
In order to study the properties of the SGWB from the stacked spectrograms, we need to reduce the contamination from the background noise.

 The realistic noise on each individual detector can be non-Gaussian, non-stationary and have glitches, the multiple detector noise will not show such behaviour at all times when an astrophysical signal is present in the detectors. Although the value of $\Delta t=0$ (that is when signals are detected with maximum correlation in two different detectors) plays a special role for the signal, it is not a special point for the detector noise. The contamination to $P_{cc}(\Delta t, \Delta f)$ due to noise will be similar at all values of $\Delta t$ due to stationarity. We can assume translation symmetry of the noise power spectrum of the detectors over the time scale over which the signal is present in the time-frequency domain spectrogram analysis. For any value of $\Delta f$, the contribution of noise to  $P_{cc}(\Delta t, \Delta f)$ can be estimated by the average (or median) value of $P_{cc}(\Delta t, \Delta f)$ for $\Delta t \neq 0$ and its neighbourhood which depends on the typical time scale of the GW signal.
 
\subsection{Signal extraction and noise statistics}
 In order to characterize the noise distribution and extract the contribution of the signal from the $\Delta t,\Delta f \simeq 0$ region we can define an aperture intensity (like aperture photometry in astronomical imaging) in the time-frequency domain as 
 
\begin{align} \label{aperture}
P^{\rm aperture}_{cc}(\Delta t_i \pm \epsilon, \Delta f) \equiv \frac{1}{2\epsilon}\sum_{x=\Delta t_i -\epsilon}^{\Delta t_i + \epsilon} P_{cc}(x, \Delta f),
\end{align}
where $\epsilon$ (i.e. the half-size of the aperture) is the time scale over which we define the aperture intensity. Here we consider $\epsilon$ to be such that the cross correlation due to the signal peaking at $(\Delta t, \Delta f) = 0$ spreads out to $\Delta t \simeq 0 \pm \epsilon$. Considering $N_t$ independent segments for $|\Delta t| > \epsilon$ and estimating  $P^{\rm aperture}_{cc}(\Delta t_i, \Delta f)$ we obtain a sample distribution of the noise cross-correlation $\bf{P^{back}}$ as 
\begin{align} \label{pback}
&{\bf P^{back}}(\Delta f) :=\{P^{\rm aperture}_{1}(\Delta t_1\pm \epsilon, \Delta f),
\nonumber\\
&P^{\rm aperture}_{2}(\Delta t_2\pm \epsilon, \Delta f), \ldots, P^{\rm aperture}_{N_t}(\Delta t_{N_t}\pm \epsilon, \Delta f)\}.
\end{align}

\begin{figure*}[ht!]
\centering
\includegraphics[width=14.9cm]{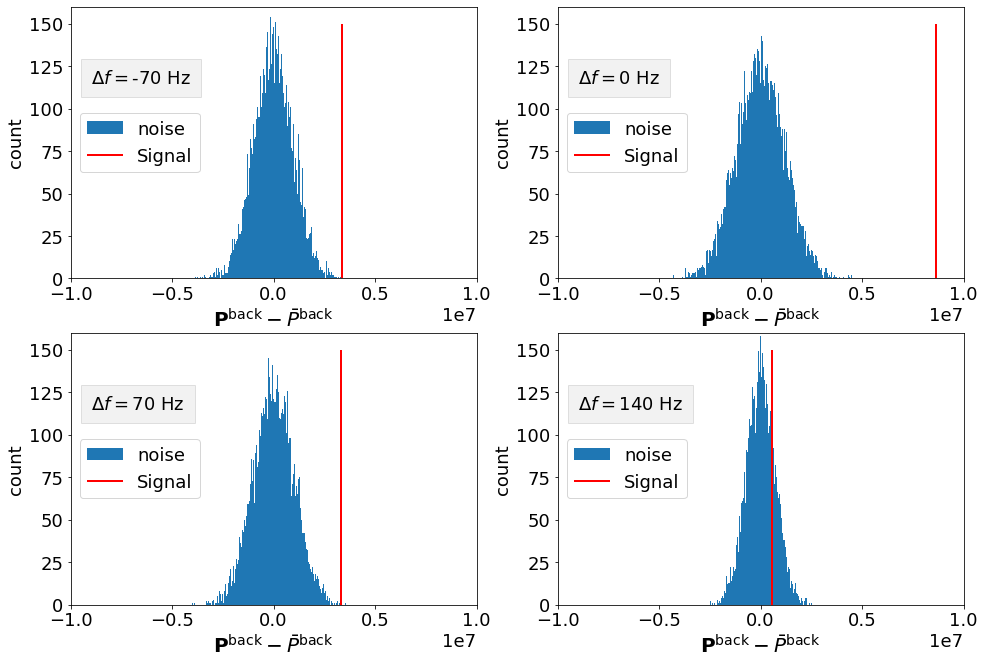}
\caption{The plot shows the mean subtracted  ${\bf P_{\rm back}}-\bar{P}^{\rm back}$ distribution as given in Eq. \eqref{pback} and Eq. \eqref{background}for different values of $\Delta f$. The red vertical lines denote the value of the \texttt{SpeCS} estimator computed for the injected signal as obtained from Dataset I at various values of $\Delta f$ using Eq. \eqref{p_sig}.} 
\label{Fig:pback_dist}
\end{figure*}

Now, to obtain the behaviour of the noise cross-correlation over the given range of $\Delta f$, we consider the average of $\bf{P^{back}}$, given as

\begin{align} \label{background}
\bar{P}^{\rm back}(\Delta f) \equiv 
\langle  \bf{P^{back}}(\Delta f) \rangle
\nonumber\\
\text{for $|\Delta t_i| > \epsilon$}.
\end{align}
The obtained mean subtracted noise distribution is shown in Fig \ref{Fig:pback_dist} at various values of $\Delta f$ (see Sec. \ref{data_gen} for more details about the simulated noise and SGWB signal used for making this plot). As can be seen, for the synthetic noise that we have used for the analysis in the paper, the distribution is approximately Gaussian. In the presence of glitches, the distribution of \textbf{P$^{\rm back}$} will be non-Gaussian. In such scenarios, one needs to estimate the noise background from the actual distribution of observed data.

 We can now define a background-cleaned unbiased estimator as
\begin{align} \label{p_clean}
P^{\rm clean}_{cc}(\Delta t,\Delta f) \equiv P_{cc}(\Delta t, \Delta f)-\bar{P}^{\rm back}(\Delta f).
\end{align}

Further, as the signal is only expected within $\Delta t=0\pm \epsilon$ we can further take the average of $P^{clean}_{cc}$ within this range to estimate a total flux:
\begin{align} \label{p_sig}
P^{\rm sig}={1\over 2 \epsilon}\sum_{\Delta t=0- \epsilon}^{0+\epsilon} P^{\rm clean}_{cc},
\end{align}
The covariance of the background noise is given as:
\begin{equation} \label{cov}
    C(\Delta f,\Delta f') \equiv {\rm Cov}[{\bf P^{back}}(\Delta f),{\bf P^{back}}(\Delta f')]
\end{equation}
The signal to noise ratio (SNR), combining the different frequency bins, can then be written as
\begin{align} \label{rho_full_tot}
\begin{split}
\rho_p&(\Delta t=0 \pm \epsilon)=\\ &\sqrt{\sum_{\Delta f\Delta f'} {\hat{P}^{\rm sig}_{cc}(\Delta f)C^{-1} (\Delta f, \Delta f') \hat{P}^{\rm sig}_{cc}(\Delta f')}}\,\,\, .
\end{split}
\end{align}
It is expected that the covariance matrix, $C(\Delta f,\Delta f')$, obtained using the LIGO detector noise will have large off-diagonal terms, and the number of independent information in each $\Delta f$ will be less. The covariance matrix  for the synthetic LIGO noise used in the paper (see Sec. \ref{data_gen} for more details) is shown in Fig. \ref{fig:cov}). The total SNR estimation can be made only within the frequency bin width ($\delta f$) around $\Delta f=0$. 


This estimator captures additional information present in the stochastic GW signal on top of the power spectrum estimation. By combining multiple pairs of GW detectors the total SNR for the detector network is given as
\begin{align} \label{rho_multi}
\rho_{\rm total}^2=\sum_{I,J>I} (\rho_{p}^{2})_{IJ}(\Delta t=0 \pm \epsilon)
\end{align}
where $I,J$ denotes all the detectors in the network.




\section{Application of SpeCS on simulated SGWB} \label{sims}
To test the efficiency and properties of the proposed estimator Eq. \eqref{speccs}, we simulate mock SGWBs by generating low amplitude (individually unresolved) black hole binary merger waveforms and injecting them in noise. We consider two distinct scenarios of SGWB based on different merger rates. 
Firstly, the merger rate is taken as relatively high and this is to show how \texttt{SpeCS} would work in an optimistic, signal-dominant case. Secondly, we consider a relatively low merger rate giving us a low density of events in order to emulate a case closer to what is expected to be observed by LVK from the low redshift merger rate. 

\subsection{Simulation setup} \label{data_gen}

\begin{figure}[ht!]
\centering
\includegraphics[width=8.5cm]{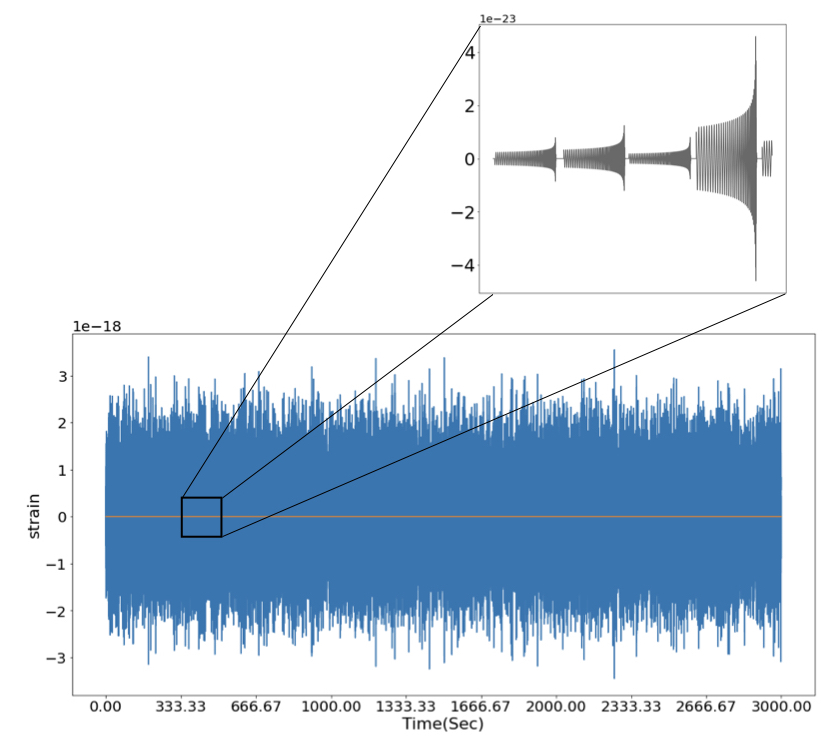}
\caption{ In this plot we show a $3 \times 10^{3}$ seconds segment of synthetic LIGO noise along with the merger waveforms injected in the noise (shown by the orange line). Subplot: The  GW waveforms generated by the IMRPhenomXAS model for dataset-I are shown. }
\label{Fig:sgwb_ts}
\end{figure}

\begin{figure}[ht!]
\centering
\includegraphics[width=8.5cm]{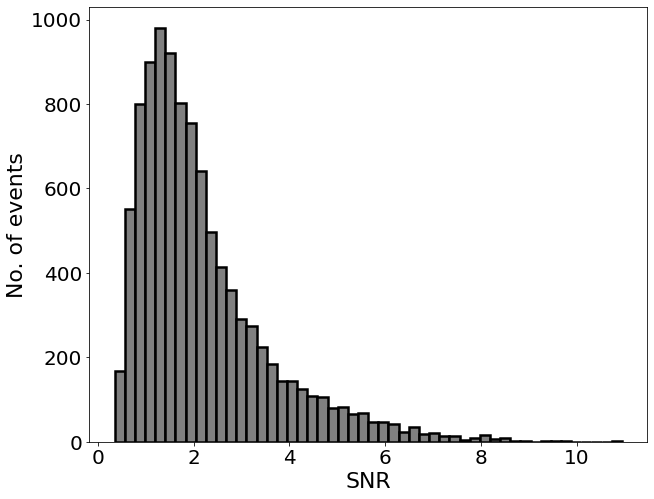}
\caption{The distribution of SNR for the individual events that were injected in the synthetic LIGO (Livingston) noise is shown. Events with SNR $< 8$ are considered sub-threshold. This SNR was computed using the PSD obtained from the O3 run of LIGO (Livingston) noise.}
\label{Fig:snr_dist}
\end{figure}

{\bf Dataset-I:}
We start with $3 \times 10^{4}$ second  segments of synthetic LIGO/VIRGO noise, generated with PyCBC \cite{alex_nitz_2022_6912865} using the PSDs of L, H and V detectors (in particular for the O3 observational run).  $10^4$ individual non-overlapping BH binary merger waveforms were generated using the IMRPhenomXAS model\cite{Pratten:2020fqn} and projected individually for the L, H and V detectors. This was done by randomly sampling the sky location parameters (from a uniform distribution) as well as taking into account the appropriate time lag between the individual signals depending on the source location. The simulated signals were then injected into the synthetic LIGO noise with randomised time spacing between consecutive events Fig.\ref{Fig:sgwb_ts}. For the simulated BH merger waveforms the masses were taken between 10-90 $M_\odot$, spin $(-0.8,0.8)$ and distance $6 \times 10^{3}$ - $3 \times 10^{4}$ Mpc. Fig. \ref{Fig:snr_dist} shows the SNR distribution of the individual waveform injections.

\begin{figure}[ht!]
\centering
\includegraphics[width=8.5cm]{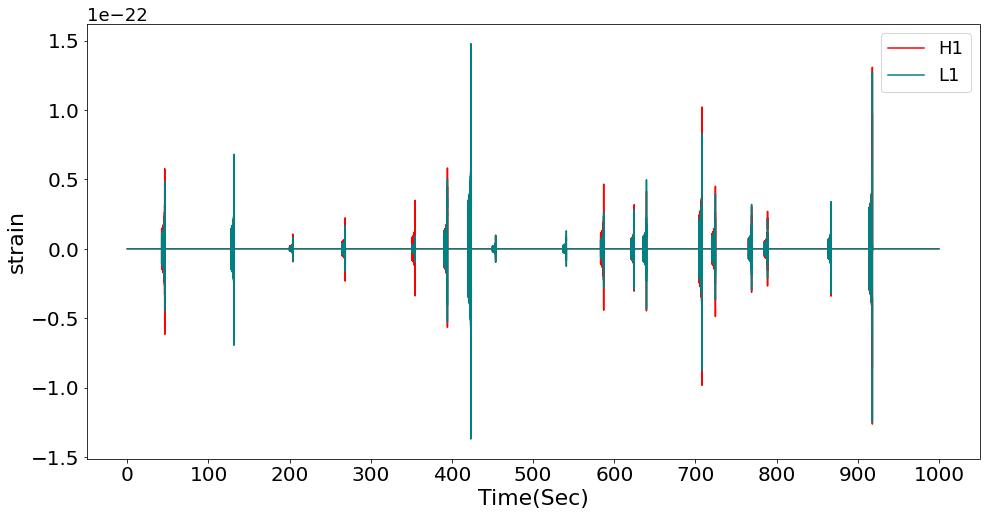}
\caption{A $10^{3}$ sec segment of the simulated stochastic GW signal (dataset-II) is shown for the L1 and H1 detectors in the absence of noise. }
\label{Fig:wave}
\end{figure}

{\bf Dataset-II:}
We start with a $5 \times 10^{5}$  second segment of L and H detector noise, generated by PyCBC using the PSD for the O3 observational run of LIGO. $10^{4}$ individual BH binary merger waveforms (with parameter range same as the case I) were injected into the synthetic detector noise after projecting the waveforms in the respective (L and H) detector frames for randomised sky location. For this dataset a significantly low merger rate is taken compared to Dataset-I. Fig. \ref{Fig:wave} shows a $10^{3}$ sec segment of the simulated waveform for L and H detectors (without the inclusion of noise).

\begin{figure}
    \centering   
    \includegraphics[width=8.5cm]{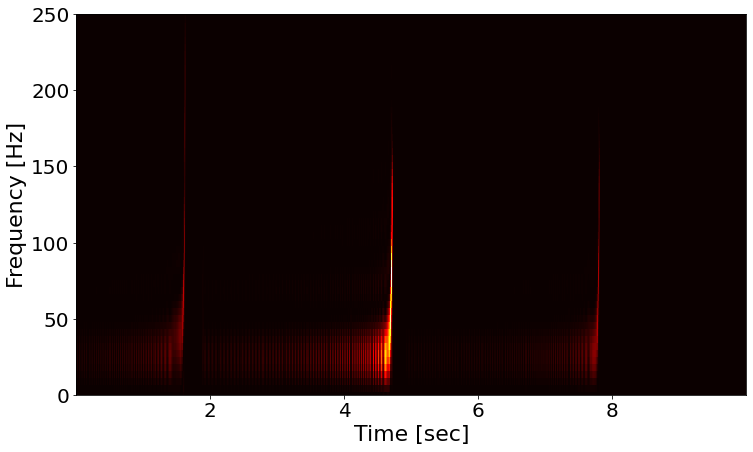}
    \includegraphics[width=8.5cm]{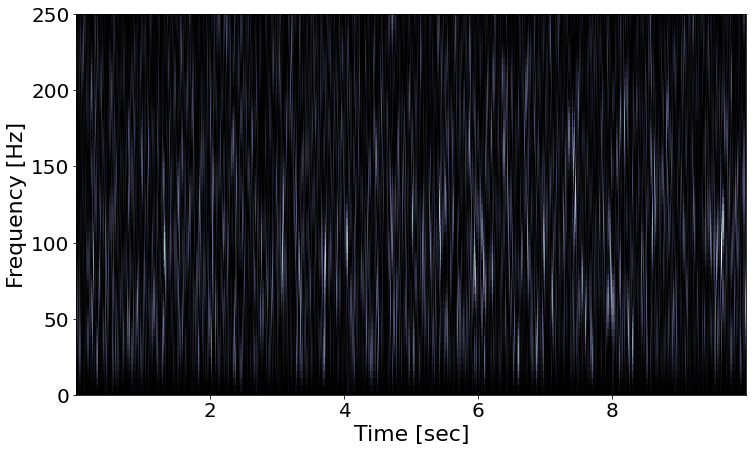}
    \caption{The spectrograms generated from the simulated GW waveforms and noise are shown here. On top, we show the waveforms generated for the L1 detector (dataset-I) and the bottom plot shows the synthetic noise generated using the PSD of the L1 detector. Here, only for illustration we showed the spectrograms till 250Hz but for our analysis, we choose $f_{\rm max}=640$Hz.}  
    \label{Fig:spect}
\end{figure}

 The simulated time series signals (having a sampling rate of 4096 Hz) were first whitened using the respective noise PSD's and then converted into spectrograms by performing a short-time-Fourier transform ( sampling rate=4096 Hz, length of each segment=78/4096 sec, number of points to overlap between segments=64/4096 sec, Tukey window). In Fig. \ref{Fig:spect} a 10 seconds segment of the spectrogram of noise and the GW waveforms is shown for dataset I. As the high frequency SGWB signal is not relevant for the analysis we cropped the spectrograms at $f_{\rm max}=640$Hz.

\begin{figure}[ht!]
\centering
\includegraphics[width=9cm]{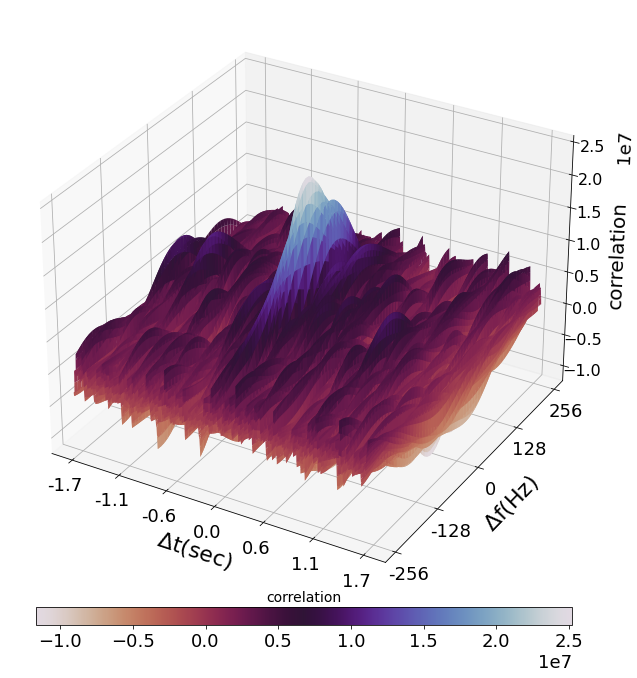}
\caption{ The cross-correlation of the spectrogram, $P_{cc}^{\rm clean}$, generated using dataset I for L and H detectors. The peak in the $(\Delta t, \Delta f)=0$ region indicates the presence of the correlated GW background signal. }
\label{Fig:LH_corr}
\end{figure}





\subsection{ \texttt{SpeCS} analysis and results}

\begin{figure}
    \centering
    \includegraphics[width=8.5cm]{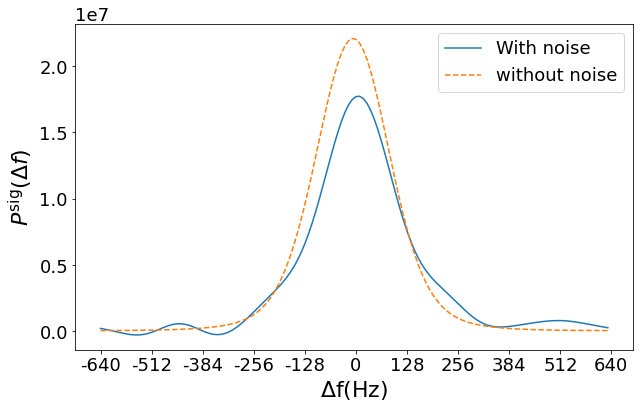}
    \caption{This plot shows the behaviour of $P^{\rm sig}$ as a function of $\Delta f$. The blue solid line is for $P^{\rm sig}$ computed using the L and H signal from Dataset I and the orange dotted line is the same quantity computed for the waveforms of Dataset I but without injecting them into noise. For the dashed orange curve (i.e the one obtained without the noise) we did not perform any background subtraction as given in Eq. \eqref{p_clean} and hence the amplitude is comparitively higher. }
    \label{fig:p_sig}
\end{figure}

In order to search for a SGWB signal, the data set (which can extend for hours, days or even years) is usually split into shorter segments. For datasets I and II we consider segments of 500 seconds and then compute $P_{cc}(\Delta t, \Delta f)$ of N such segments (denoted by $P_{cc}^i$).  Let ${\bf p}:=\{P_{cc}^1, P_{cc}^2,\ldots, P_{cc}^N\}$ be a set of measurements obtained by computing the cross-correlation of the $N$ statistically independent fragments of the dataset. The mean of this stack is given as
\begin{align} \label{s_mean}
    \langle{\bf p}\rangle={1\over N}\sum_{i=1}^{N} P^{i}_{cc}.
\end{align}
This mean of the \texttt{SpeCS} estimator, as computed for the SGWB signal, would contain noise contamination as the signal is noise dominated. Following Eq. \eqref{p_clean}, we can perform the background subtraction on $\langle{\bf p}\rangle$ in order to get the unbiased estimator for the correlated GW signal (as shown in Fig \ref{Fig:LH_corr} for dataset I).
Further for Dataset I, $P^{\rm sig}$ as given in Eq. \eqref{p_sig} is shown in Fig \ref{fig:p_sig}. 

\begin{figure}
    \centering
    \includegraphics[width=8.5cm]{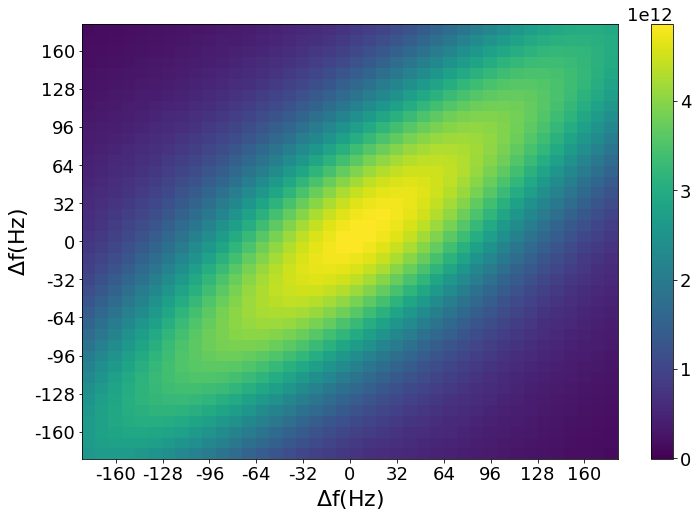}
    \caption{The covariance matrix of the background noise obtained using Eq. \eqref{cov}. For this illustration we choose $|\Delta f|<180$Hz. }
    \label{fig:cov}
\end{figure}

\begin{figure}[ht!]
\centering
\includegraphics[width=8.5cm]{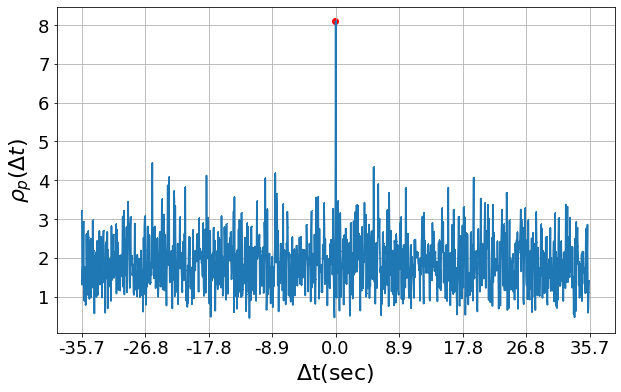}
\includegraphics[width=8.5cm]{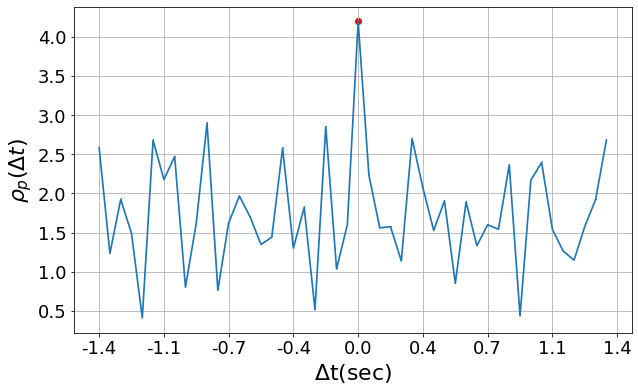}
\caption{ Plot of $\rho_p$ obtained for dataset I (top) and II(bottom) by computing $P^{\rm sig}$ at various different values of $\Delta t$.}
\label{Fig:sample_snr_t}
\end{figure}

 For both dataset I and II, after computing the corresponding $P^{\rm sig}$ we use Eq. \eqref{cov} to compute the covariance as shown in Fig. \ref{fig:cov}.
 In order to finally compute the SNR we use the obtained $P^{\rm sig}$ and $C(\Delta f, \Delta f')$ in Eq. \eqref{rho_full_tot}. In Fig \ref{Fig:sample_snr_t} the SNR obtained by choosing the aperture at different values of $\Delta t$ is shown.  For this computation, we choose $\epsilon=$0.02 seconds (i.e 12 $\Delta t$ bins) and perform the $\Delta f$ summation from $\Delta f=+18$Hz to $\Delta f=-18$Hz (i.e 4 $\Delta f$ bins centered around $\Delta f=0$).  We observed that a specific choice of these parameters does not affect the SNR significantly unless we select a value that cuts off the peak in the correlation (Fig \ref{Fig:LH_corr}) abruptly.
 We add both positive and negative $\Delta f$ bins to get the final SNR as $P^{\rm sig}$ contains unique information for positive and negative values of $\Delta f$. This is due to the fact that the \texttt{SpeCS} estimator cross-correlates signal from two different detectors having independent detector frequencies for the signal. Further, from Fig \ref{fig:cov} it is evident from the small values of the covariance matrix at $\{+\Delta f,-\Delta f\}$ that the information content is unique at positive and negative $\Delta f$.
 The averaged peak value at $|\Delta t|< 0\pm \epsilon$ for Dataset I and II denotes the strength of the signal. The values at $|\Delta t|> 0\pm \epsilon$ denote the contribution from the noise. For Dataset I and II we obtained a peak SNR of 8.1 and 4.2 respectively.
 
\begin{figure}[ht!]
    \centering
    \includegraphics[width=8.5cm]{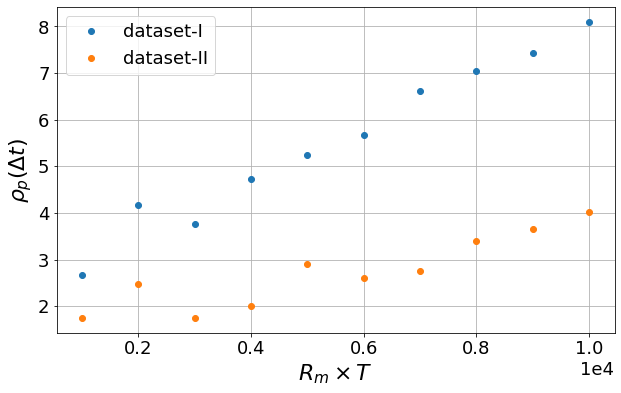}
    \caption{$\rho_{p}(\Delta t)$ obtained for dataset I and II considering various length ($T$) of the datasets. }
    \label{fig:snr_inj}
\end{figure}

 In dataset-II, even though the effective merger rate was taken to be much lower compared to dataset-I, the \texttt{SpeCS} method was able to efficiently isolate the signal from the dominant background noise. Fig \ref{fig:snr_inj} shows how $\rho_P$ builds up with an increase in the total time of integration for both Datasets I and II. Also, as shown in Appendix \ref{multi_det} the overall detection SNR increases while combining signals from multiple detectors as commonly done by the LVK collaboration.

\subsection{Comparison with standard cross-correlation estimator}

In order to study and compare the efficiency of the \texttt{SpeCS} method with the standard optimal cross-correlation estimator Eq. \eqref{cf} we compute the SNR of dataset I using both methods. For this comparison, we compute the SNR using only the GW waveforms from dataset I without injecting them into the synthetic detector noise. This was done in order to remove the influence of different noise realisations on the SNR enabling us to infer the factor by which \texttt{SpeCS} improves the SNR for the given dataset.  One must note that in order to obtain the SNR the variance (Eq. \eqref{sigma}) and the covariance matrices (Eq. \eqref{cov})   were computed for the respective synthetic detector noise. 

For the standard cross-correlation method, we consider a set of maximum frequencies $\{ f^{\rm max}_1,f^{\rm max}_2,...,f^{\rm max}_i\}$ upto which the summation in Eq.\eqref{C_hat} was performed for each element of $\{ f^{\rm max}_i \}$ followed by computing the corresponding $\sigma(f^{\rm max}_i)$ each time and then finally obtaining $\text{SNR}_{CC}(f^{\rm max}_i)$ using Eq. \eqref{cc_snr} for the entire set of $f^{\rm max}_i$. 

\begin{figure}
    \centering
    \includegraphics[width=8.5cm]{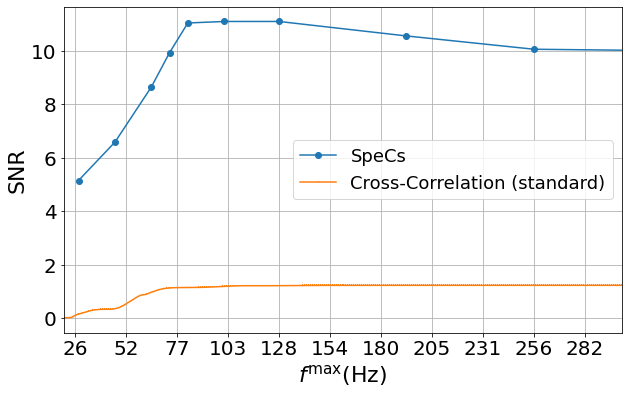}
    \caption{This figure compares the SNR obtained using the \texttt{SpeCS} method as proposed in this paper with the standard optimal cross-correlation SNR \cite{Allen:1997ad} for different $f^{\rm max}$ frequencies. Here we computed the SNR using just the GW waveforms, and without injecting them into noise (except the variance was computed using noise). This is done in order to make the estimator explicitly unbiased so that any bias/fluctuations in the SNR due to different noise realizations are removed and we get the factor by which \texttt{SpeCS} improves the detectability of the SGWB signal as compared to the standard method. The  \texttt{SpeCS} SNR shows a peak at $f^{\rm max} \sim 100$ Hz as the noise contamination increases at higher frequencies where the signal is not dominantly present (see text for discussion).}   
    \label{fig:Cfdiff}
\end{figure}

For the \texttt{SpeCS} method, firstly, we compute $P_{cc}(f^{\rm max}_i)$ as given in Eq. \eqref{speccs} by using the spectrogram, $S_{s_i}(t,f^{\rm max}_i)$, calculated upto the $f^{\rm max}_i$ frequencies. Using this $P_{cc}(f^{\rm max}_i)$ we calculate $\rho_{p(f^{\rm max})}$ for each element in $\{ f^{\rm max}_1,f^{\rm max}_2,...,f^{\rm max}_i\}$. As shown in Fig. \ref{fig:Cfdiff} this enables us to study how SNR builds up as a function of frequency till it saturates at some high value of $f^{\rm max}$. For dataset-I using the noiseless signal, the SNR saturates at 1.2 while using the standard CC method and it saturates at 9.8 using \texttt{SpeCS}. From this, we can conclude that the \texttt{SpeCS} method performs $\approx 8$ times better in detecting the SGWB signal when compared to the standard CC method.
 Given that \texttt{SpeCS} is model-agnostic, it is not an optimal estimator for parameters of any particular model. Therefore, as seen in Fig. \ref{fig:Cfdiff}, its SNR can decrease due to contamination by noisy data as we increase $f^{\rm max}$. In future work, we will explore the performance of \texttt{SpeCS} for different models and explore how we can achieve optimal measurement of the signal depending on the noise covariance matrix.

\section{Discussion and conclusion}\label{conc}


In this paper, we proposed a novel method, Spectrogram correlated stacking (or \texttt{SpeCS}), to constrain/detect the time-dependent astrophysical stochastic gravitational wave background originating from non-overlapping binary sources, having a low duty cycle.  Firstly, we obtained the spectrograms of the GW signal by mapping the time-domain GW detector strain to the time-frequency domain using a Short-Time-Fourier-Transform. Then the \texttt{SpeCS} method is implemented which is based on computing the time-frequency domain cross-correlation between the spectrograms. 

The main advantage of \texttt{SpeCS} is that since it is a higher-order detection statistics, it probes  the non-gaussian time-dependent aspect of the signal. Since the GW signal is a chirp in the time-frequency domain, SpeCS would take into account correlations present between different frequency bins of the signal thus probing the excess information hidden in the chirping nature of the non-overlapping GW events. Further, the time-frequency correlation between various detectors helps to distinguish it from the background noise. The noise properties of each detector would be very particular and it won't exhibit such a high degree of time or frequency correlations thus enabling us to detect/extract the GW signal which would remain the same (apart from a phase difference) across multiple detectors. These features make the higher-order \texttt{SpeCS} method more efficient in extracting information from the SGWB signal and complement the more traditional cross-correlation method used by LVK \cite{Allen:1997ad, LIGOScientific:2021psn}. 
We show that the SNR obtained using the \texttt{SpeCS} method is $\approx 8$ times higher than the SNR obtained using the standard cross-correlation technique Fig. \ref{fig:Cfdiff}. For this comparison, we computed the SNR of Dataset I using the \texttt{SpeCS} method and the standard cross-correlation method but without injecting the waveforms into the corresponding detector noise.

In this paper, we mainly focused on introducing the \texttt{SpeCS} formalism, and thus its efficiency was tested using a simulated SGWB dataset generated using synthetic LIGO noise(based on the PSD of the O3 run\cite{list2020sensitivity}). For the real-world application of this method on the time-series signal as obtained/detected by LVK, the presence of glitches would be a concern. Glitches and more general non-stationary noise would influence the overall value of the cross-correlation and interfere with the underlying SGWB signal. Glitches, for example, can be dealt with by applying gating where they are removed from the time-series data by using a suitable window \cite{KAGRA:2021kbb}, but we defer a detailed study of the impact of non-stationary noise on \texttt{SpeCS} to future work. Another possible systematic is the correlated noise in-between detectors, e.g., caused by the Schumann resonances in the Earth's magnetosphere, which must be modeled alongside with Astrophysical SGWB contribution to \texttt{SpeCS} \cite{Romano:2016dpx,Janssens:2022tdj,Himemoto:2023keu}. Furthermore, in an ongoing study, we are exploring the applicability of the \texttt{SpeCS} to perform parameter estimation \cite{mandic2012parameter,Romano:2016dpx} and extract information about the BH population and event rates. From the tilt/slope of the correlation function in the time-frequency domain one can possibly break the degeneracy amongst mass distribution, the merger rate, and other astrophysical parameters while performing the parameter estimation of the SGWB.

\section*{Acknowledgment }
The authors are thankful to Jishnu Suresh for reviewing the manuscript and providing useful comments. LFLM thanks the financial support of the São Paulo Research Foundation(FAPESP) grant 2021/09531-5 under the BEPE program, thanks Pennsylvania State University for the hospitality during the realization of this project and knowledges funding from the EU Horizon under ERC Consolidator Grant,
no. InspiReM-101043372. The work of SM is a part of the $\langle \texttt{data|theory}\rangle$ \texttt{Universe-Lab} which is supported by the TIFR and the Department of Atomic Energy, Government of India. NA is funded by the University of Waterloo, the National Science and Engineering Research Council of Canada (NSERC) and the Perimeter Institute for Theoretical Physics. Research at Perimeter Institute is supported by the Government of Canada through Industry Canada and by the Province of Ontario through the Ministry of Economic Development \& Innovation.
The authors would like to thank the LIGO-Virgo-KAGRA Scientific Collaboration for providing the noise curves. 
This research has made use of data or software obtained from the Gravitational Wave Open Science Center (gw-openscience.org), a service of LIGO Laboratory, the LIGO Scientific Collaboration, the Virgo Collaboration, and KAGRA. LIGO Laboratory and Advanced LIGO are funded by the United States National Science Foundation (NSF) as well as the Science and Technology Facilities Council (STFC) of the United Kingdom, the Max-Planck-Society (MPS), and the State of Niedersachsen/Germany for support of the construction of Advanced LIGO and construction and operation of the GEO600 detector. Additional support for Advanced LIGO was provided by the Australian Research Council. Virgo is funded, through the European Gravitational Observatory (EGO), by the French Centre National de Recherche Scientifique (CNRS), the Italian Istituto Nazionale di Fisica Nucleare (INFN) and the Dutch Nikhef, with contributions by institutions from Belgium, Germany, Greece, Hungary, Ireland, Japan, Monaco, Poland, Portugal, Spain. The construction and operation of KAGRA are funded by Ministry of Education, Culture, Sports, Science and Technology (MEXT), and Japan Society for the Promotion of Science (JSPS), National Research Foundation (NRF) and Ministry of Science and ICT (MSIT) in Korea, Academia Sinica (AS) and the Ministry of Science and Technology (MoST) in Taiwan. This material is based upon work supported by NSF's LIGO Laboratory which is a major facility fully funded by the National Science
Foundation. 
\appendix
\section{Spectrogram resolution and SNR}
\begin{figure}[t!]
\centering
\includegraphics[width=8.5cm]{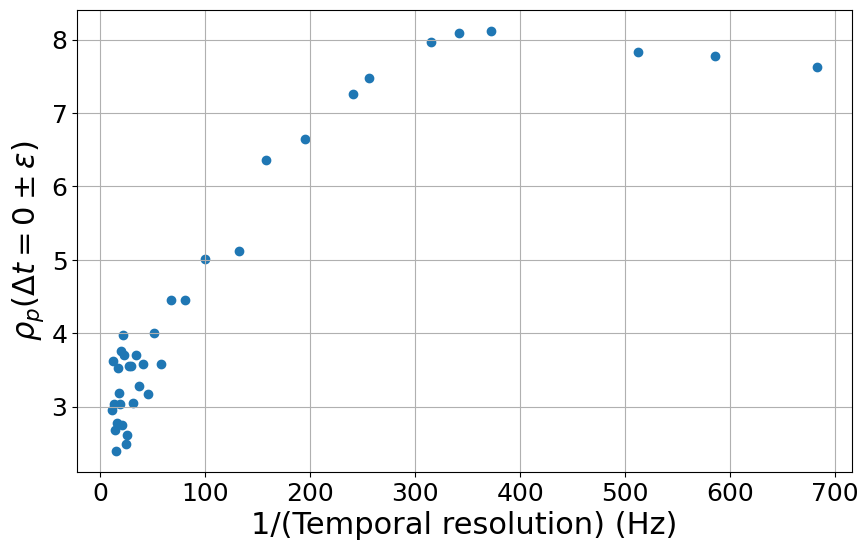}
\caption{Top: The plot shows the variation of the maximum value of $\rho_p$ at $|\Delta t| <\epsilon$ with respect to the different temporal resolutions of the spectrogram.The temporal resolution is the window size used for generating the spectrogram.
}
\label{Fig:snr_res_sub}
\end{figure}

The time and frequency resolution of the spectrogram (determined by the sampling rate of the signal, the width of the window function, and overlap) can strongly influence the efficiency of the \texttt{SpeCS} method. If we choose a time resolution that is extremely high/fine the signal will get over-resolved, and random noise fluctuations in the spectrogram will start contributing spuriously to the cross-correlation. On the other hand, if the time resolution is not high enough the waveforms won't be resolved adequately and their contribution to the cross-correlation would be blurred by the noise. The optimal situation is to be in the intermediate regime where the signal is resolved properly and the \texttt{SpeCS} estimator is not overly sensitive to small changes in the resolution. To establish the robustness of the proposed \texttt{SpeCS} estimator to variations in resolution, we perform the cross-correlation analysis for a range of time resolutions (obtained by varying the width of the window and the overlap) and present the results in Fig. \ref{Fig:snr_res_sub}. There is a plateau in the time resolution axis where the peak value of $\rho(\Delta t)$ remains stable and for the analysis done in this paper we choose the parameters of the spectrogram in such a way that the resolution lies well within this stable regime. 


\section{Three detector analysis} \label{multi_det}

The optimal strategy for the detection of SGWB involves performing a cross-correlation analysis with the signal detected by multiple detectors and then computing the combined SNR \cite{Meacher:2015iua}.

\begin{figure}[ht!]
    \centering
    \includegraphics[width=8.5cm]{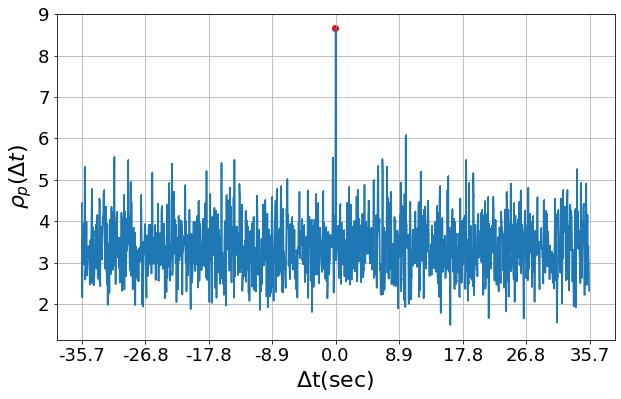}
    \caption{$\rho_{\rm total}$ obtained for the LVH detector network by combining baselines as given in Eq. \eqref{rho_multi}  }
    \label{fig:3_det}
\end{figure}

We use the simulated signals obtained for all three detectors in dataset I and implement the \texttt{SpeCS} method on the spectrograms of the signal obtained for three possible combinations of detector pairs (LH, LV, HV).  
For the three detectors, the combined SNR $\rho_{\rm total}=8.7$  \eqref{rho_multi} is shown in Fig. \ref{fig:3_det}.

\section{Standard cross-correlation analysis}\label{usuallcc}

In this section, we compare the efficiency of \texttt{SpeCS} to the conventional cross-correlation based optimal search techniques used by LIGO. Given two GW detectors a cross-correlation statistic $C(f)$ can be defined as:
\begin{align} \label{cf}
    C(f)={2\over T} {\text{Re} [s_I^*(f)s_J(f)]\over \gamma_T(f) S_0(f)} ,
\end{align} 
where $T$ is the total duration of the signal, $s_{(I,J)}$ is the Fourier transform of the strain for each detector that we are considering, $S_0(f)$ is the spectral shape of the expected GW background,

\begin{align}
    S_0(f)={3 H_0^2\over 10\pi^2f^3},
\end{align}
 and $\gamma_T(f)$ is the overlap reduction function \cite{Allen:1997ad,LIGOScientific:2021psn}. 


In dataset-I since the waveform injections are made into a definitive realisation of the detector noise we can compute $C(f)$ for noise+SGWB as well as for just the signal. 

The variance of the estimator $C(f)$, in the absence of correlated noise and in the small SNR limit is given as
\begin{align}
   \sigma^2(f) = \frac{1}{2T \Delta f} \frac{P_I(f) P_J(f)}{\gamma^2_{IJ} (f) S^2_0(f)},
\end{align}
where T is the observation time, $\gamma$ is the overlap reduction function, $P_I$ and $P_J$ are the noise power spectrum for each detector and $\Delta f$ is the frequency resolution.

 Using $C(f)$ as given in Eq. \eqref{cf} one can construct an optimal estimator \cite{Romano:2016dpx} where the contribution from various frequency bins are combined and given as
 \begin{align} \label{C_hat}
     \hat{C}={\sum_k C(f_k)\sigma^{-2}(f_k)\over \sum_k \sigma^{-2}(f_k)}.
 \end{align}
The variance of this optimal estimator is given as 
\begin{align} \label{sigma}
    \sigma=\bigg[\sum_f \sigma^{-2}(f)\bigg]^{-1/2}.
\end{align}
The SNR is given as
\begin{align} \label{cc_snr}
    \rm{SNR}_{cc}={\hat{C}\over \sigma}.
\end{align}

We compute the $\rm{SNR}_{CC}$ for dataset-I using \eqref{cc_snr} by diving the data stream into segments of 192 s which are Hann-windowed having an overlap of 50\%, then compute the discrete Fourier transform of each segment and resample the resulting spectrum to a  frequency resolution of 1/32 Hz.
Fig. \ref{fig:Cfdiff} shows the cumulative $SNR$ as a function of frequency. For dataset-I we get a maximum SNR of 1.2 when using just the waveforms without injecting them into noise.

\bibliography{sgwb}

\end{document}